\documentclass[
               biblatex,     
               ]{jacow}

\usepackage[english]{babel}

\begin{document}

\title{THE CANTED COSINE THETA HTS SEXTUPOLE DEMONSTRATOR OF FCC-EE}

\author{M.~Koratzinos\thanks{michael.koratzinos@cern.ch}, F.~Bardi, V.~Batsari, I.~Dimoulios, O.~Kuhlmann, A.~Thabuis\\ European Organization for Nuclear Research, Geneva, Switzerland\\ 
M.~Duda, Paul Scherrer Institute, Villigen, Switzerland}

\maketitle


\begin{abstract}
A single-aperture, two-layer Canted-Cosine-Theta (CCT) sextupole magnet using high-temperature superconducting (HTS) ReBCO tape has been developed for the short straight sections (SSS) of FCC through the FCCee-HTS4 project \cite{HTS4}.  The magnet was designed, manufactured and tested under cryogenic conditions. Two HTS tapes from two manufacturers have been qualified for this specific application. Design and manufacturing details and cryogenic temperature measurements are presented. This demonstrator represents the first HTS CCT magnet ever constructed.
\end{abstract}

\section{INTRODUCTION}

The design of the Future Circular Collider (FCC-ee) aims to increase luminosity while reducing operational costs and environmental impact. In the current baseline \cite{benedikt_future_2025}, arc magnets are based on normal-conducting technology, resulting in significant power consumption. Sextupoles contribute a substantial fraction of this load and are therefore a key target for innovation.

The FCCee-HTS4 project investigates the use of high-temperature superconducting (HTS) magnets in the collider arcs. By reducing electrical losses and enabling more compact nested magnet layouts, this approach can improve machine performance and reduce RF requirements \cite{HTS4_CHART_2025}.

A \qty{240}{mm} long HTS canted cosine theta (CCT) sextupole demonstrator was designed and constructed using insulated HTS tape conductors. The CCT design was selected for its excellent stress management, high field quality, and simple manufacturing. Aluminium formers with CNC-machined grooves accommodate the tape stacks, and the coils are wound manually without tape pre-tensioning.

\section{MATERIALS AND METHODS}

\subsection{Design and manufacturing}

\begin{figure}
    \centering
    \includegraphics[width=0.9\linewidth]{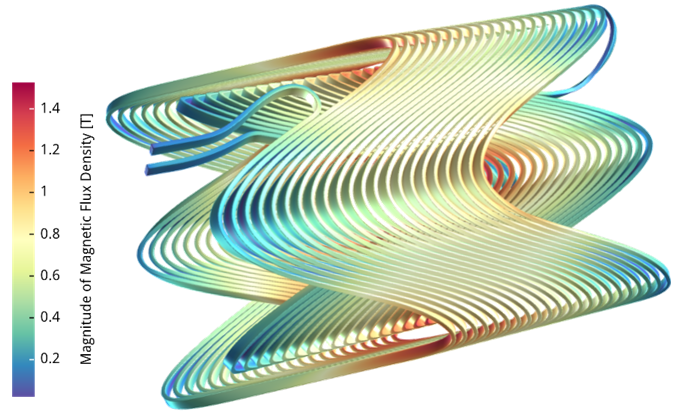}
    \caption{Conductor trajectory colored by the magnitude of the magnetic flux density (T). }
    \label{fig:condPath}
\end{figure}

The target magnetic field parameters are summarized in Table~\ref{tab:field_parameters}.  The maximum operating current is kept low to reduce heating losses through the current leads. The  canted cosine theta (CCT) sextupole geometry follows a Frenet-Serret trajectory which ensures no hard-way bending \cite{frenetserret}. It has been computed using the RAT suite of programs \cite{RATGUI}. Computed field quality is excellent, helped by the large aperture. The optimized conductor path features a minimum bending radius of \qty{4.85}{mm} and a groove width of \qty{1.6}{mm}, corresponding to the accommodation of a stack of \qty{10}{} HTS tapes. In total, approximately \qty{300}{m} of HTS conductor is needed for the winding of the demonstrator. The selected groove geometry represents a compromise between electromagnetic performance and manufacturability (in particular it follows a binormal Frenet-serret geometry), while remaining compatible with the mechanical limits of the conductor and the practical constraints of winding. The conductor path, coloured by magnetic the magnitude of the flux density at the operating current of \qty{260}{A}, is shown in Fig.~\ref{fig:condPath}. The magnet was tested without iron shielding. Iron shielding, according to simulations, can increase the field gradient by up to \qty{20}{\%}, depending on its distance from the coil.

\begin{table}[!hbt]
   \centering
   \caption{Main design parameters of the HTS CCT sextupole demonstrator using the double-sided \qty{4}{mm} wide HTS tape from Faraday Factory Japan.}
   \begin{tabular}{lc}
       \toprule
       \textbf{Specifications} & \textbf{Value} \\
       \midrule
       Magnet type               & CCT             \\
       Aperture                  & \SI{90}{mm}     \\
       Operating current         & \SI{260}{A}     \\
       Operating temperature     & \SI{40}{K}      \\
       Field gradient            & \SI{1000}{T/m^2}\\
       Max. field at conductor   & \SI{1.5}{T}     \\
       Inductance                & \SI{12.6}{mH}     \\
       Critical current fraction & \SI{23}{\percent}\\
       Temperature margin        & \SI{27}{K}\\
       \bottomrule
   \end{tabular}
   \label{tab:field_parameters}
\end{table}

Because the grooves are not perpendicular to the former axis and follow a continuously varying three-dimensional trajectory, the winding former was manufactured using 5-axis CNC machining. Al~6082-T6 was selected as the former material owing to its favorable cryogenic properties, including good mechanical strength, dimensional stability, and thermal conductivity at low temperature. X-ray computed tomography (CT) was used to verify the geometric conformity of the machined former. This non-destructive technique enabled three-dimensional reconstruction of the groove geometry and winding trajectory, allowing direct comparison with the nominal CAD model and identification of local geometric deviations or placement errors. The results were excellent. The RMS of deviation of manufacturing accuracies  for the outer former is \qty{40}{\micro m}.

The curvature of the winding path subjects the conductor to alternating tensile and compressive strain depending on the local orientation along the trajectory. To accommodate these mechanical constraints, a commercially available double-sided \qty{4}{mm} wide HTS tape supplied by Faraday Factory Japan LLC was selected, in which two coated conductors are laminated in a face-to-face configuration with their superconducting layers facing each other. This geometry reduces the distance of the superconducting layer from the neutral axis, thereby lowering the strain experienced during bending. Each tape consists of a YBCO layer deposited on a \qty{40}{\micro m} Hastelloy substrate, with silver stabilisation layers (\qty{2}{\micro m} on the HTS side and \qty{1}{\micro m} on the substrate side), copper stabilisation layers of \qty{5}{\micro m} per side, and a polyimide insulation coating. The conductor was qualified at \qty{77}{K} prior to winding, exhibiting no degradation of critical current (\qty{320}{A}) for bending radii down to \qty{3.5}{mm}, confirming its suitability for the demonstrator geometry. A second HTS conductor supplied by Shanghai Superconductor Technology (SST) was also qualified to extend the mechanical margin. The tape has a \qty{30}{\micro m} Hastelloy substrate, with a \qty{5}{\micro m} copper surround and a \qty{55}{\micro m} copper lamination on the HTS side, and is insulated with a \qty{12.5}{\micro m} Kapton wrapping, resulting in a total thickness of approximately \qty{160}{\micro m}. This conductor showed no measurable degradation of critical current (\qty{180}{A}) down to bending radii of \qty{2.5}{mm}, demonstrating improved tolerance to tight winding conditions.


\subsection{Winding}
A stack of \qty{10}{} tapes was bundled using a custom-made PTFE guiding tool and manually inserted into the CCT groove. The winding path is IP-protected and was used under licence. The tape insulation was found to be insufficient, as after a few turns most tapes became electrically shorted to the aluminium formers. At the end of the winding process, only two tapes remained electrically insulated.

\begin{figure}
    \centering
    \includegraphics[width=0.8\linewidth]{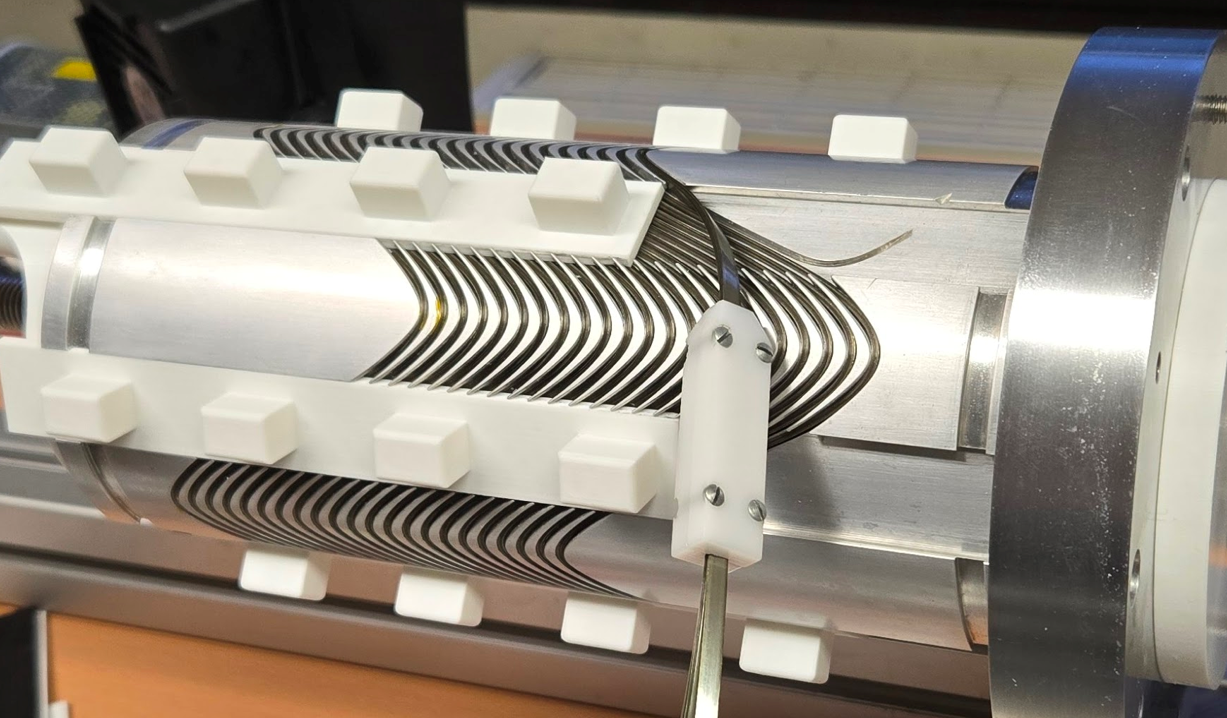}
    \caption{Outer former winding process showing insertion of the \qty{10}{}-tape stack into the groove using a custom-made PTFE guiding tool.}
    \label{fig:winding}
\end{figure}

\subsection{Impregnation}

After coil assembly, the sextupole demonstrator was impregnated with paraffin wax to provide mechanical support and thermal contact \cite{impregnation-wax}. As wax shrinks by \qty{15}{\%} when it solidifies, a special jig was used.
The system comprised a vacuum filling circuit and a temperature-controlled setup with water-cooled Peltier elements. The magnet was clamped between copper plates, their temperature regulated by the Peltier elements. A vacuum pump reduced internal pressure to \qty{1.1}{mbar} before molten wax was injected. 

Directional solidification was achieved by imposing a vertical temperature gradient: the bottom plate was progressively cooled while the top remained warmer. This minimized void formation, while shrinkage during solidification was compensated by continuous wax supply from above. The process continued with decreasing temperatures until full solidification.
Operating up to \qty{90}{\degreeCelsius} with a maximum gradient of \qty{36}{\degreeCelsius}, this method provides a simple, reversible impregnation technique. Inspection confirmed excellent wax packing with negligible voids; \qty{750}{\g} of wax were used.

\subsection{HTS Tape Splicing}
The \qty{10}{} tapes were spliced in pairs, resulting in a total of nine splice joints of \qty{150}{\mm} length. The joints were fabricated using a custom-made aluminum splicing press with integrated heaters, PT100 sensors, and spring-loaded clamps to ensure uniform temperature and pressure during solder reflow. The applied pressure was estimated at approximately \qty{3.5}{\MPa} \cite{splices-bala}.

Prior to splicing, the polyimide insulation varnish was chemically removed using a \qty{10}{\%} NaOH solution. The tapes were then aligned and joined using Sn63/Pb37 no-clean solder paste (TS391AX10). A controlled thermal cycle with a peak temperature of \qty{205}{\degreeCelsius} and total heating time below \qty{5}{min} was used to ensure proper solder reflow while limiting thermal degradation of the HTS conductor. 

\begin{figure}
    \centering
    \includegraphics[width=0.9\linewidth]{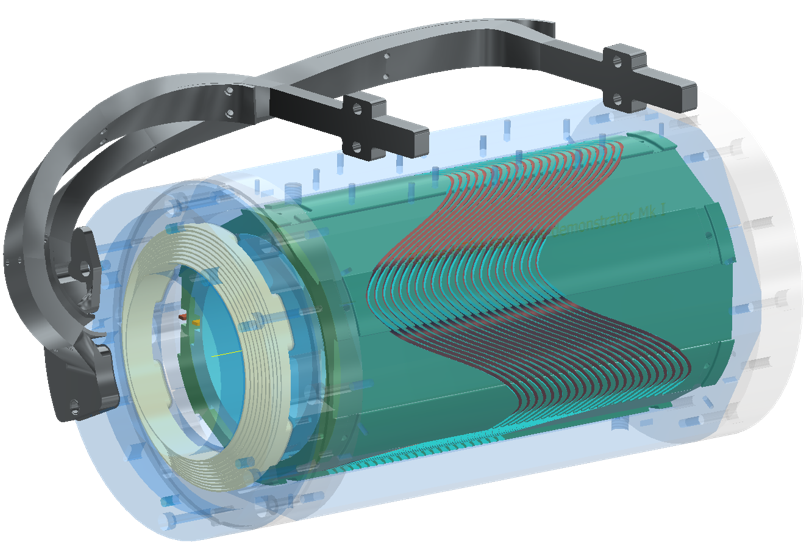}
    \caption{Magnet assembly highlighting structural components; the splice box, current lead supports, and outer former are shown opaque for clarity }
    \label{fig:assembly}
\end{figure}

The completed joints were insulated with Kapton tape and inserted into the circular grooves of the splice box, which is visible in Fig.~\ref{fig:assembly}, with inner and outer radii of \qty{50}{mm} and \qty{65}{mm}, respectively. Previous tests showed that bending to radii down to \qty{45}{mm} did not increase splice resistance or degrade the critical current.


\subsection{Current leads}

In order to avoid hot spots in the in and out HTS tapes, an aluminium support closely matching the path of the tape (3D printed) provided thermalization to the aluminium magnet sleeve.  

\section{TESTING AT COLD}
\begin{figure}
    \centering
    \includegraphics[width=0.8\linewidth]{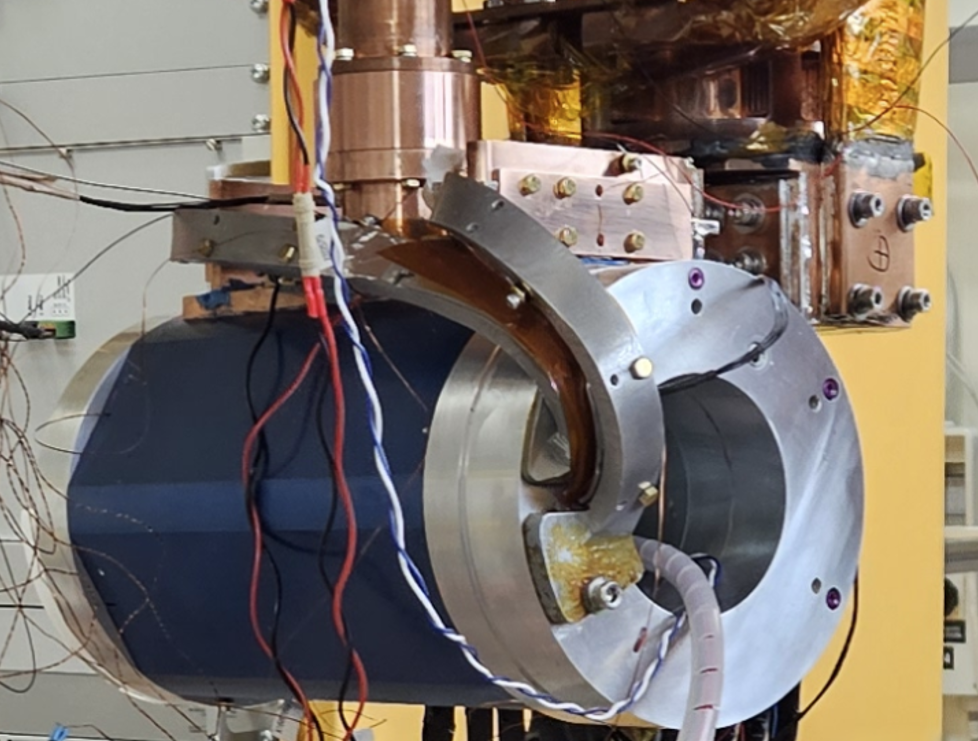}
    \caption{The magnet connected to the cryogen-free test stand.}
    \label{fig:cryostat}
\end{figure}

\begin{figure}
    \centering
    \includegraphics[width=1\linewidth]{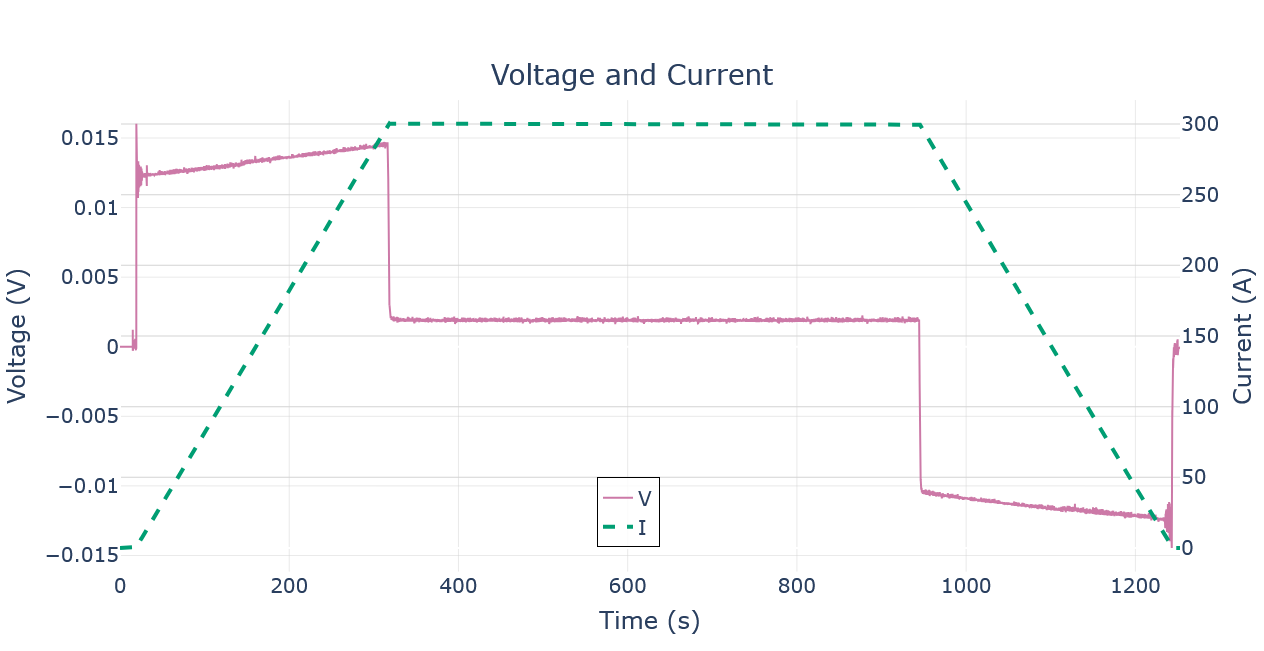}
    \caption{Current and voltage curves vs time during the second \qty{300}{A} ramp.}
    \label{fig:VandI}
\end{figure}

\begin{figure}
    \centering
    \includegraphics[width=1\linewidth]{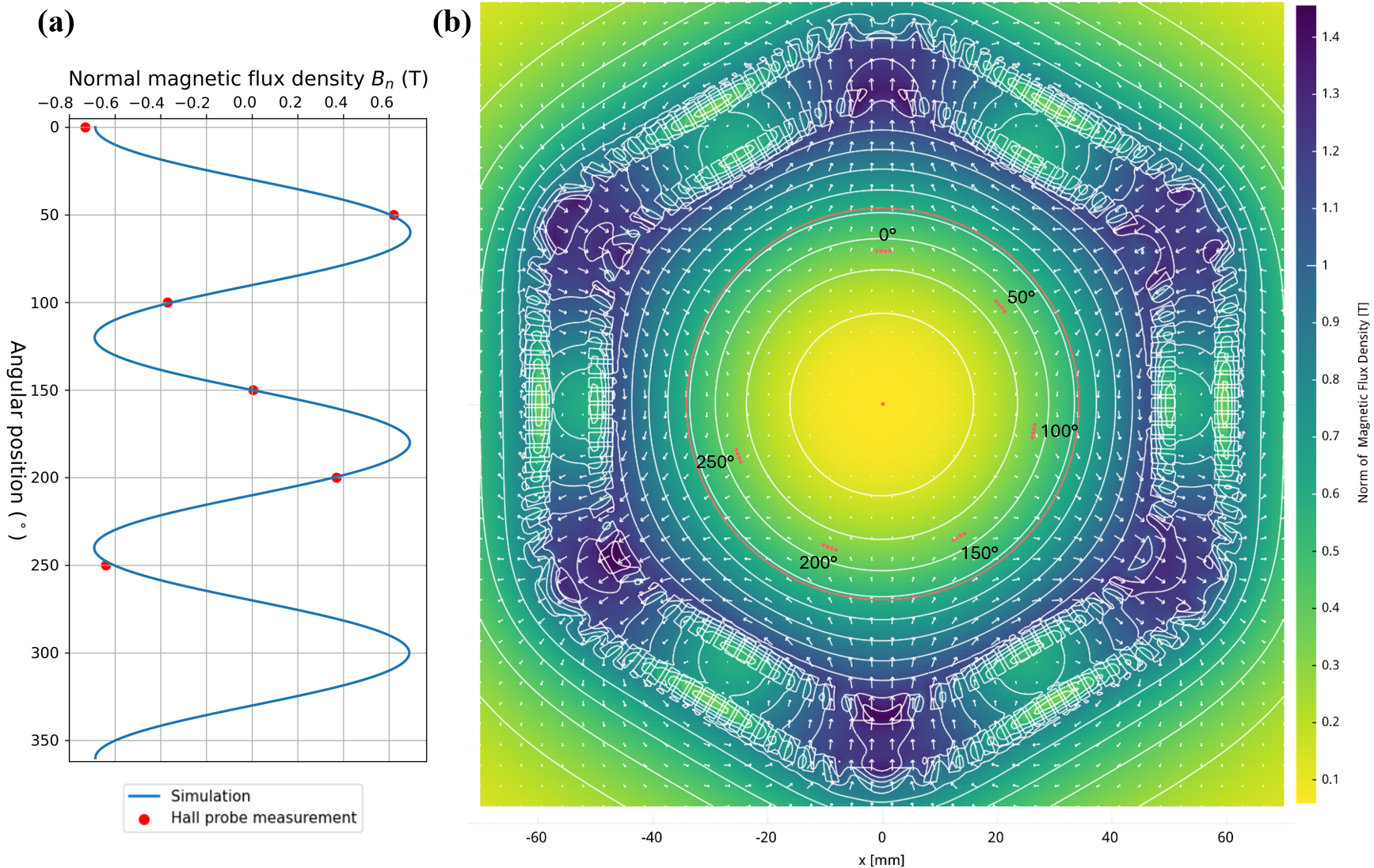}
    \caption{
(a) Simulated and measured normal magnetic flux density, $B_n$, at a radius of \qty{35}{mm}.
(b) Heat map of $B_n$ with Hall probe positions indicated in red. Both results are shown for a transport current of \qty{300}{A}.
}
    \label{fig:hallsensors}
\end{figure}

The magnet was tested in a cryogen-free test stand using an RDE-418D4 \qty{4}{K} cryocooler head, as shown in Fig.~\ref{fig:cryostat}.
The cooldown took approximately \qty{17}{h} and reached a minimum temperature of \qty{11}{K}. Temperature monitoring was performed using five Cernox CU sensors located on the bottom of the magnet, the copper current leads, and the aluminium supports. Two \qty{20}{\ohm} resistive heaters installed near the cryocooler connection were used for temperature regulation. With a constant heating power of \qty{21}{W}, a stable operating temperature of \qty{46.5}{K} was achieved. Magnetic field measurements were performed using five Toshiba GaAs THS119 Hall sensors mounted on a circular PCB at the magnet center, at a radius of \qty{35}{mm}, and distributed at angular positions of \qtylist{0;50;100;150;200;250}{\degree}. The transport current was measured using a DCCT sensor. Magnet protection was based on a total voltage threshold of \qty{15}{mV}; upon triggering, the current supply ramped down at a rate of \qty{10}{A/s}. This protection scheme was adopted due to the damaged insulation in the conductor stack, preventing the development of high-voltage arcs during a quench event. The current was ramped at \qty{1}{A/s} up to \qty{300}{A}. Following an initial stepped excitation in increments of \qty{20}{A}, the magnet was energized to \qty{300}{A} and held for \qty{600}{s}, during which stable voltages were observed, as shown in Fig.~\ref{fig:VandI}. At \qty{300}{A} a total voltage of \qty{1.9}{mV} was measured, attributed to splice resistances with an average value of \qty{700}{n\ohm}, corresponding to a dissipated power of \qty{0.57}{W}. During the current hold, an average temperature increase of \qty{0.4}{K} was observed. A peak magnetic field magnitude of \qty{0.73}{T} was measured at the \qty{0}{\degree} probe position. An average field drift of \qty{3}{mT/h} was observed across the Hall sensors. A comparison of the normal component of the magnetic flux density between Hall probe measurements and simulation results is shown in Fig.~\ref{fig:hallsensors}(a), alongside (b) the computed heat map, where the Hall probe locations are highlighted.

\section{CONCLUSION}

The world’s first HTS CCT sextupole magnet has been designed, constructed, and successfully tested. The magnet demonstrated stable operation beyond its nominal conditions in terms of both transport current and operating temperature. A very good agreement was observed between measured magnetic fields and numerical simulations, validating the design and modeling approach. Future work will focus on quantifying the temperature margin and performing detailed field quality measurements.

A second sextupole magnet with more demanding specifications will be wound and tested for crab cavity applications to further assess the robustness of the technology. The SST tape with Kapton insulation will be used, as varnish-based electrical insulation proved insufficient during winding.

\section{ACKNOWLEDGEMENTS}

 This work was done under the auspices of the CHART programme of FCC. Use of proprietary winding technology under license is acknowledged.

\printbibliography

\end{document}